\documentclass[12pt,a4paper]{article}
\usepackage[a4paper,rmargin=1.7cm,lmargin=1.7cm,tmargin=1.0cm,bmargin=1.0cm,includefoot]{geometry}

\usepackage{array}
\usepackage{epsfig}

\usepackage{cite}

\usepackage[namelimits]{amsmath}
\usepackage{amssymb}
\usepackage{amsfonts}
\usepackage{bbm}

\usepackage{multirow}
\usepackage{color}

\usepackage[T1]{fontenc}      
\usepackage{ae}               
\usepackage{indentfirst}

\usepackage[affil-it]{authblk}

\date{}

\begin{document}

\markboth{}{\textit{D.A. Fagundes, M.J. Menon, P.V.R.G. Silva}}

\title{\bf Asymptotic Scenarios for the Proton's Central Opacity:
      An Empirical Study}

\author[1]{D.A. Fagundes}
\author[2]{M.J. Menon}
\author[2]{P.V.R.G. Silva}
\affil[1]{\small Instituto de F\'{\i}sica Te\'orica, Universidade Estadual Paulista  \\
01140-070 - S\~ao Paulo, SP, Brazil}
\affil[2]{\small Instituto de F\'{\i}sica Gleb Wataghin, Universidade Estadual de Campinas \\
13083-859 Campinas, SP, Brazil}

\maketitle

\begin{abstract}
We present a model-independent analysis of the experimental data on the ratio $X$ between the elastic 
and total cross-sections from $pp$ and $\bar{p}p$ scattering in the c.m. energy interval 5 GeV - 8 TeV. 
Using a novel empirical parametrization for that ratio as a function of the energy and based on 
theoretical and empirical arguments, we investigate three distinct asymptotic scenarios: either 
the black-disk (BD) limit or scenarios above and below that limit. Our analysis favors a scenario 
below the BD, with asymptotic ratio $X = 0.36 \pm 0.08$. 
\end{abstract}


\vspace{0.5cm}

\centerline{\textit{Presented at Diffraction 2014, Primo\v{s}ten, Croatia, September 10 - 16, 2014}}

\vspace{0.5cm}

\section{Introduction}
\label{s1}

The dependence of the ratio between the elastic and total hadronic cross-sections as a function of
the c.m. energy,
\begin{equation}
X(s) = \frac{\sigma_{el}}{\sigma_{tot}}(s),
\label{1}
\end{equation}
constitutes an important quantity in the investigation of  elastic and soft diffractive
processes. Besides giving information on the hadron's central opacity (profile function
at $b$ = 0) and on the ratio of the inelastic to total cross-sections, it is also connected with 
the ratio between the total cross 
section and the elastic slope parameter through the approximated relation
$X = \sigma_{tot}/16\pi B_{el}$.

Presently, in the lack of a theoretical framework able to describe
the elastic scattering states from the first principles of QCD, one possible way to look
for new phenomenological insights and/or inputs is the empirical approach.
In this context, Fagundes and Menon have recently developed a model-independent analysis
of the experimental data on the ratio $X$ from $pp$ scattering in the energy interval
10 GeV - 7 TeV \cite{fm}. The empirical parametrization
is given by $X(s) = A f(s)$, with $f(s) = \tanh\{a + b \ln(s/s_0) + c\ln^2(s/s_0)\}$,
where $s_0 =$ 1 GeV$^2$, $a, b, c$ are dimensionless free fit parameters and
$A$  the asymptotic limit.
In order to estimate the uncertainties in extrapolations to higher energies, two
asymptotic limits have been considered: either $A$ = 1/2 (black-disk limit) or $A$ = 1 (maximum
unitarity). Beyond consistent data reductions of the experimental information on $X(s)$, the approximate relation
has allowed extrapolations of the uncertainty regions
in the ratio $\sigma_{tot}/B_{el}$ that may be useful in the determination
of the proton-proton total cross-section from proton-air production cross-section in cosmic-ray experiments
\cite{fm}.

In this communication, this empirical analysis of the $X$ data is updated and developed in several aspects.
The experimental data from $\bar{p}p$ scattering, all the $pp$ TOTEM data at 7 TeV (four points) and 8 TeV (one point) are 
included in the dataset and the energy cutoff is down to 5 GeV. The description of the change of
curvature in $X(s)$ demands a novel empirical ansatz for $f(s)$ and as explained in what follows,
we investigate all the three possible asymptotic scenarios: either the black-disk limit or
scenarios above or below that limit. Our main conclusions are: a) the black-disk does not represent a
definitive solution; b) the data reductions, using the novel parametrization, favor a scenario below the
black-disk, with asymptotic ratio $A = 0.36 \pm 0.08$.

After discussing the arguments for investigating the three scenarios, 
we introduce the new parametrization, discuss the fit procedures and results
and then present a summary and our
conclusions.

\section{Asymptotic Scenarios}
\label{s2}

The \textit{Black-Disk} limit represents a standard phenomenological
expectation, typical, for example, of eikonal models.
We have the arguments that follows for investigating scenarios either below or
above that limit.

\textit{Below the Black Disk.} We have 
recently developed an amplitude analysis on the quantities
$\sigma_{tot}$, $\rho$ parameter and $\sigma_{el}$, including the TOTEM
Collaboration results at 7 and 8 TeV \cite{fms,ms1,ms2}. For our purposes, we recall that the parametrization
for the total cross section is expressed by
$\sigma_{tot}(s) = \mathrm{Regge}\ \mathrm{terms}\ + \alpha + \beta \ln^{\gamma}(s/s_h)$
and fits to $\sigma_{tot}$ and $\rho$ data from
$pp$ and $\bar{p}p$ scattering above 5 GeV, led to statistically consistent solutions
either for $\gamma = 2$ (fixed) or  $\gamma > 2$ (free fit parameter).
In both cases, extension of the parametrization to $\sigma_{el}$ data (same $\gamma$ value) 
allowed to extract the ratio $X(s)$ and its asymptotic value $A$. In all cases investigated, 
we have obtained $A < 1/2$ within the uncertainties and lowest central value around $0.3$ 
(see a summary of the results in \cite{ms2}, Figure 10).
Moreover, we recall that in the publications by the TOTEM Collaboration, the authors quote the 
COMPETE Collaboration prediction for $\sigma_{tot}(s)$ \cite{compete}, presenting also
their own fit to the $\sigma_{el}(s)$ data \cite{totem}. As shown in \cite{ms2},
from these two results and using the central values of the parameters, one obtains
$X(s) \rightarrow A = 0.436$ as $s \rightarrow \infty$, suggesting, therefore
a scenario below the black disk (see also this point in \cite{ms2}, Figure 10).

\textit{Above the Black Disk.}
As discussed
in \cite{fm}, besides the obvious maximum bound allowed by
Unitarity, namely $A = 1$, the
U-matrix unitarization scheme by Troshin and Tyurin
predicts an asymptotic limit beyond the black disk, $1/2 < A \leq 1$ \cite{tt}.
Here we also recall that
in a formal context, two well known bounds for the total and inelastic cross-sections read
\cite{bounds}: 
\begin{eqnarray}
\sigma_{tot}(s) < \frac{\pi}{m_{\pi}^2} \ln^2(s/s_0),
\qquad
\sigma_{inel}(s) < \frac{\pi}{4m_{\pi}^2} \ln^2(s/s_0).
\nonumber
\end{eqnarray}
Therefore, in case of simultaneous saturation of both bounds
as $s \rightarrow \infty$, it is possible that 
$\sigma_{inel}/\sigma_{tot} \rightarrow 1/4$, which from unitarity, 
implies in $X(s) \rightarrow A = 3/4 = 0.75$.

\section{Parametrization, Fit Procedures and Results}
\label{s3}

Our dataset comprises all the experimental data on the ratio $X$ from $pp$ and $\bar{p}p$
scattering in the energy interval from 5 GeV up to 8 TeV (41 points, 28 from $pp$
and 13 from $\bar{p}p$) \cite{pdg}. With this enlarged set (as compared
with that in \cite{fm}), preliminary tests led us to change the parametrization
used in \cite{fm} by the following suitable empirical ansatz
\begin{eqnarray}
f(s) = \tanh\{\alpha + \beta \sqrt{\ln(s/s_0)} + \gamma\ln(s/s_0)\},
\label{e2}
\end{eqnarray}
where $s_0 = 25$ GeV$^2$ (the energy cutoff), $\alpha, \beta$ and $\gamma$ are free fit parameters
and for $A$ representing the asymptotic limit, the ratio is given by
\begin{eqnarray}
X(s) = A f(s).
\label{e3}
\end{eqnarray}

The data reductions have been performed with the objects of the class TMinuit of ROOT Framework, with confidence
level fixed at 68 \%. For tests on the goodness of fit we shall consider the reduced Chi squared,
$\chi^2/\nu$, and the corresponding integrated probability, $P(\chi^2, \nu)$.
Since the parametrization is non-linear in three parameters, different initial values  must be tested
in order to check the stability of the result. 
We have considered two variants in the fit procedures: either $A$ fixed, so as to \textit{impose}
an asymptotic limit, or $A$ as a free fit parameter, in order to \textit{select} a possible asymptotic
scenario.

\textit{Variant 1 - $A$ Fixed.}
We have developed 5 tests with the three scenarios:
(1) below the black-disk, either $A = 0.3$ (lowest value we have obtained in \cite{fms,ms1,ms2})
or $A = 0.436$ (the result from the TOTEM and COMPETE parameterizations);
(2) the black disk, $A = 0.5$; (3) above the black-disk, either $A=0.75$ (possible ``formal" result) 
or $A=1$ (maximum unitarity).
The statistical information on the fit results are given in Table \ref{t1}
and the comparison with the experimental data in Figure \ref{1}. As illustration it is also shown
the estimation of the ratio $X$ from the Pierre Auger Collaboration results for $\sigma_{tot}$
and $\sigma_{inel}$ at 57 TeV (not included in the dataset).
We conclude that all results
present consistent and equivalent descriptions of the experimental data analyzed. In other
words, the fit results with our empirical parametrization can not discriminate or select an asymptotic scenario.

\begin{table}[ht]
\centering
\caption{\label{t1} Statistical information on the fit results with Variant 1,
$\nu$ = 38 DOF.}
\vspace{0.2cm}
\begin{tabular}{cccccc}\hline\hline
$A$ (fixed):     & 0.3 & 0.436 & 0.5& 0.75 & 1.0 \\ \hline
$\chi^2/\nu$:     & 0.789 & 0.774 & 0.778 & 0.787 & 0.790 \\
$P(\chi^2, \nu)$: & 0.812 & 0.840 & 0.843 & 0.823 & 0.818 \\
\hline
 \end{tabular}
\end{table}

\begin{figure}[ht]
 \centering
\epsfig{file=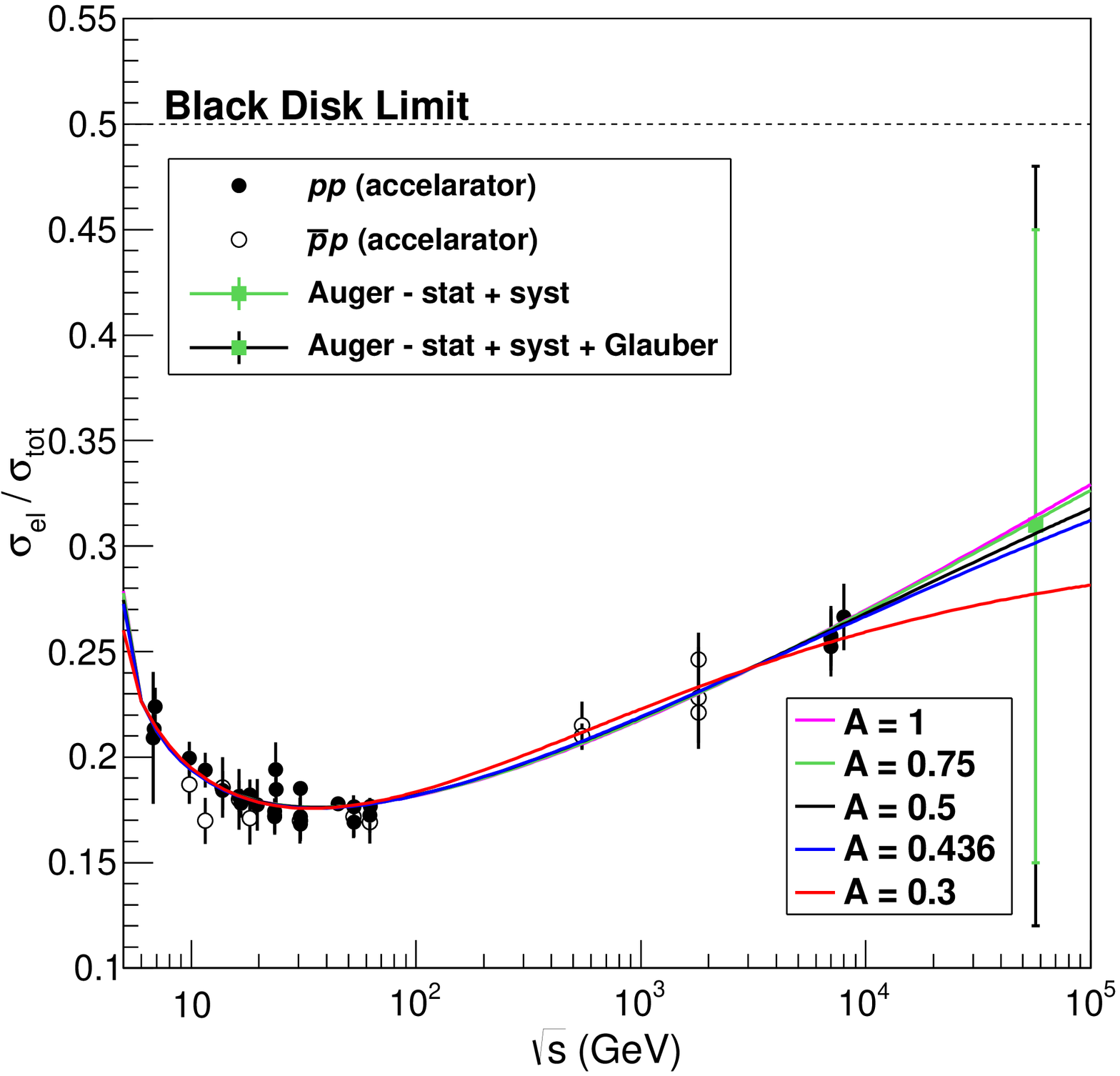,width=8.7cm,height=9.5cm}
\epsfig{file=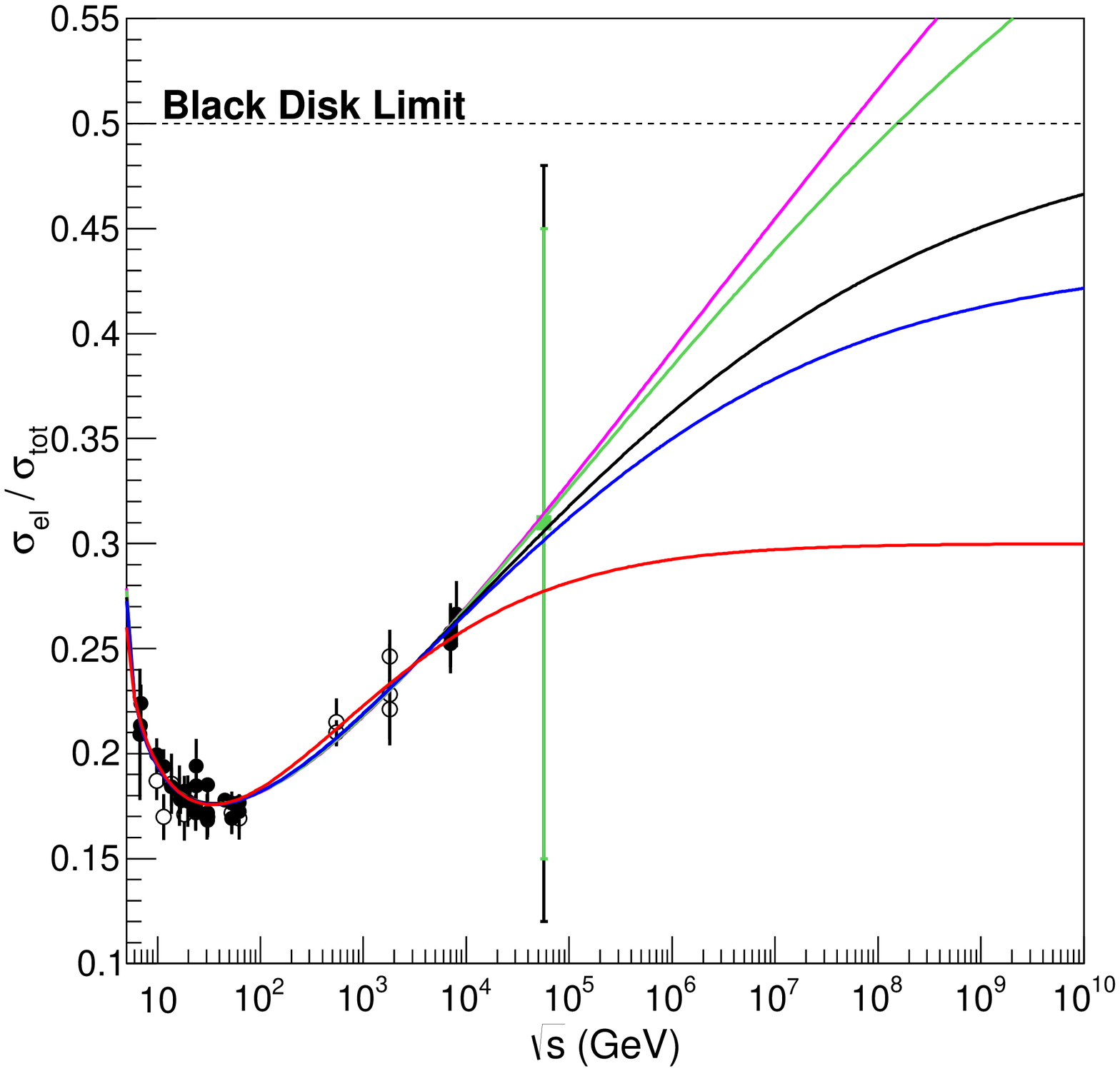,width=8.7cm,height=9.5cm}
 \caption{\label{f1} Fit results for the ratio $X(s)$ with Variant 1 ($A$ fixed) and experimental data:
up to $\sqrt{s} = 100$ TeV (left) and extrapolation to higher energies (right).}
\end{figure}

\begin{figure}[ht]
 \centering
\epsfig{file=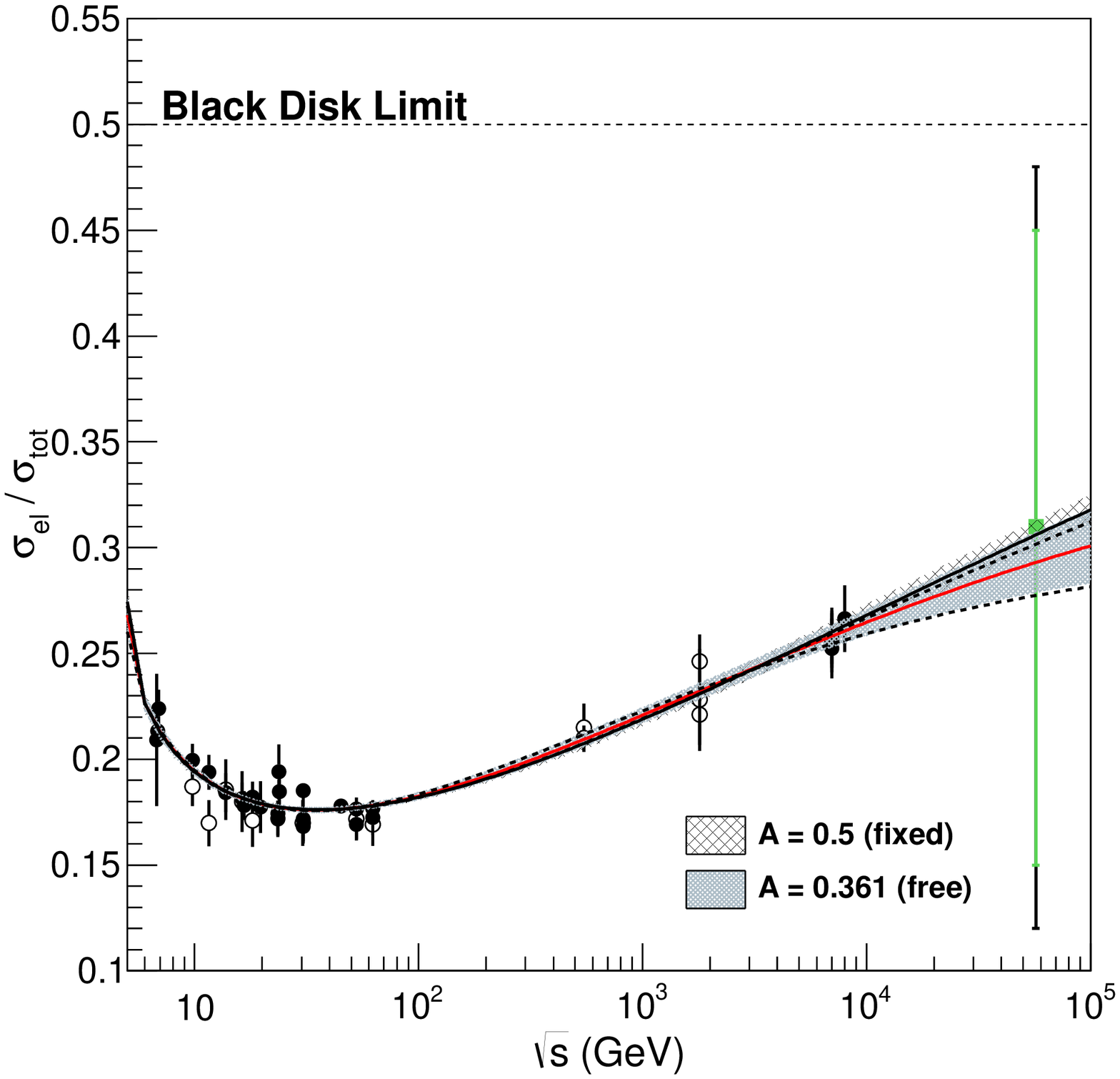,width=8.7cm,height=9.5cm}
\epsfig{file=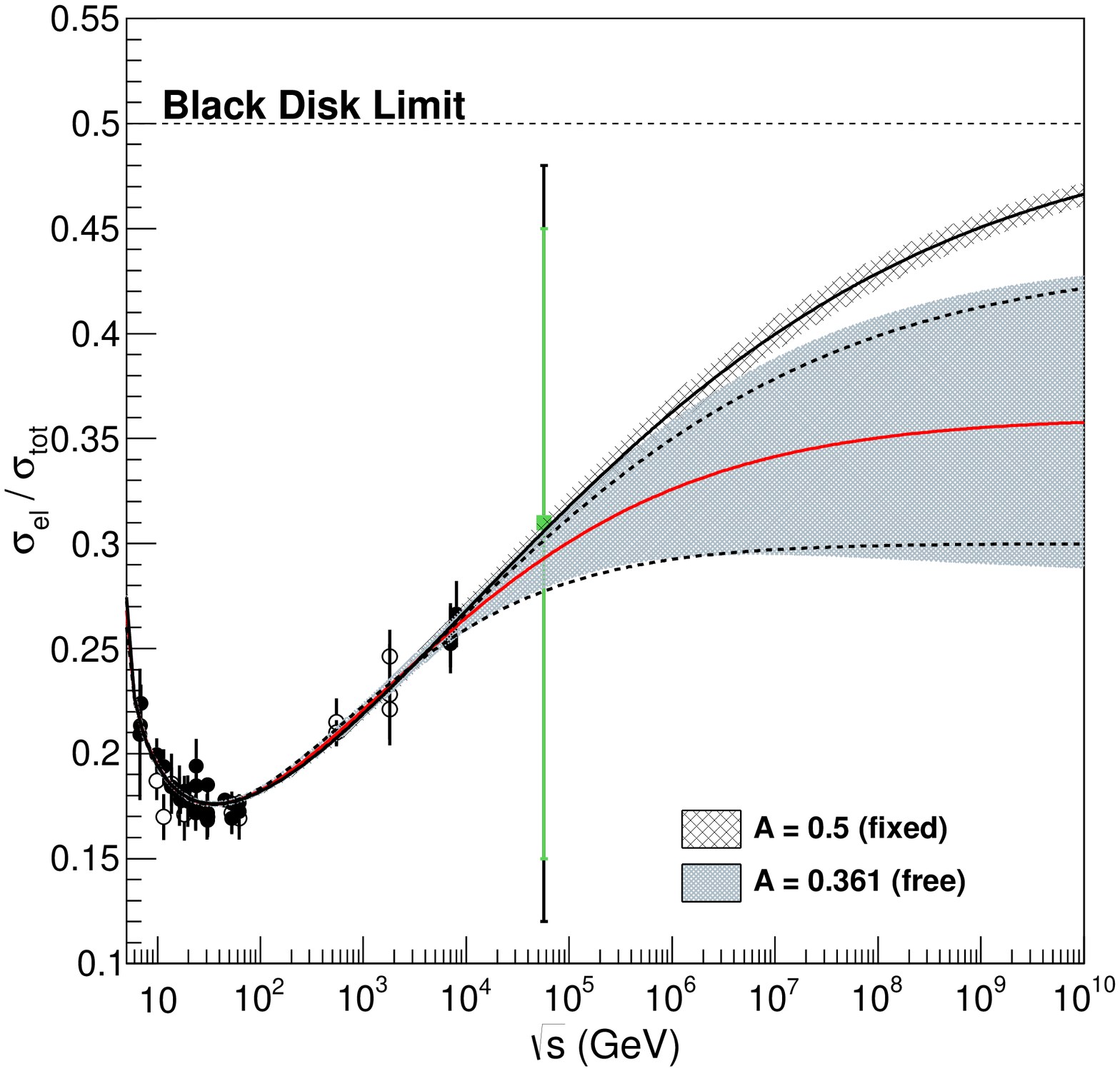,width=8.7cm,height=9.5cm}
 \caption{\label{f2} Fit results for the ratio $X(s)$ with Variant 2 ($A$ as a free fit parameter) and experimental data:
up to $\sqrt{s} = 100$ TeV (left) and extrapolation to higher energies (right).}
\end{figure}

\textit{Variant 2 - $A$ as a Free Parameter.}
Using as initial values of the parameters the final values obtained in 
Variant 1 and the corresponding values of $A$, the data reductions with
four free parameters lead to the selected asymptotic scenario, defined by
the final value of $A$.
In the 5 cases investigated the data reductions converged to an unique solution
within the uncertainties in the free parameters, with statistical results
$\chi^2/\nu$ = 0.791 and $P(\chi^2, \nu)$ = 0.814, for $\nu$ = 37 DOF and
the following values of the free parameter:
\begin{eqnarray}
A = 0.361 \pm 0.078,\ \alpha = 0.96 \pm 0.32,\
\beta = - 0.43 \pm 0.19,\ \gamma = 0.109 \pm 0.048.
\nonumber
\end{eqnarray}
The fit result with the uncertainty region, evaluated through analytical error propagation from
the free parameters (one standard deviation), is displayed in Figure \ref{f2}. For comparison, we have also
included the result and corresponding uncertainty region for the case $A = 0.5$ fixed (black disk) and
the central values for the cases $A = 0.436$ and $A = 0.3$ (same as Figure \ref{1}).
We conclude that, asymptotically and within the uncertainties, our solution is not compatible
with the black disk limit and the central values for the cases $A = 0.75$
and $A = 1.0$ neither (namely scenarios above the black disk). The central values in the cases
of $A = 0.3$ and $A = 0.436$ lie within our uncertainty region.
Therefore, our unique solution favors a scenario below the black-disk limit
and we can infer, also from unitarity: 
\begin{eqnarray}
\frac{\sigma_{el}}{\sigma_{tot}} \rightarrow 0.36 \pm 0.08, \quad
\frac{\sigma_{inel}}{\sigma_{tot}} \rightarrow 0.64 \pm 0.08 \quad \mathrm{as} \quad s \rightarrow \infty.
\nonumber
\end{eqnarray}

\section{Summary and Conclusions}
\label{s4}

We have introduced a novel suitable analytical parametrization for the ratio $X$ 
and developed two variants as fit procedures to our dataset ($pp$ and $\bar{p}p$ data
at 5 GeV $\leq \sqrt{s} \leq$ 8 TeV).
In Variant 1, we impose different asymptotic limits by fixing $A$ at 0.3, 0.436, 0.5, 0.75 and
1.0. All the results are consistent with the experimental data (Table 1 and Figure 1).
Although the results do not discriminate an asymptotic scenario
we can conclude that the black disk limit does not represent
a definitive or unique solution. 
In Variant 2, with $A$ as a free parameter, we have obtained an unique convergent solution,
indicating a scenario below the black-disk: $A = 0.36 \pm 0.08$. Within the uncertainty, this
asymptotic value is in agreement with the results obtained by Fagundes, Menon and Silva 
\cite{fms,ms1,ms2}, the prediction from the parameterizations
by the COMPETE and TOTEM Collaborations \cite{compete,totem} and also with
recent phenomenological analysis by Kohara, Ferreira and Kodama which indicates
$A$ approximately $1/3$ \cite{kfk}.

As we have discussed \cite{fms,ms1,ms2}, a scenario below the black disk is not in disagreement
with the Pumplin bound, namely
\begin{eqnarray}
\frac{\sigma_{el}}{\sigma_{tot}} + \frac{\sigma_{diff}}{\sigma_{tot}} \leq 1/2,
\nonumber
\end{eqnarray}
where $\sigma_{diff}$ stands for the soft diffractive processes (single and double dissociation).
Therefore, in case of  saturation of the Pumplin bound, it is possible to infer
$\sigma_{diff}/\sigma_{tot} \rightarrow 0.14 \pm 0.08$ as $s \rightarrow \infty.$

\section*{Acknowledgments}

Research supported by FAPESP-CAPES, Contract 2014/00337-8 (D.A.F), CNPq and FAPESP, Contract 2013/27060-3 (P.V.R.G.S.).

\end{document}